%% file: main.tex
\documentclass[nofootinbib,
 twocolumn,
 amsmath,amssymb,
 aps,
hyperref,
 prd,
 showpacs,
 superscriptaddress,
]{revtex4-1}

\usepackage{mathrsfs}
\usepackage{xspace}
\usepackage{gensymb}
\usepackage{comment}
\usepackage{graphicx}
\usepackage{dcolumn}
\usepackage{bm}
\usepackage{color}
\usepackage{xcolor}
\usepackage[colorlinks]{hyperref}

\usepackage[normalem]{ ulem }
\usepackage{soul}

\def\spi{INTEGRAL/SPI\xspace}
\def\xmm{XMM-Newton\xspace}
\def\extp{eXTP\xspace}
\def\ths{THESEUS\xspace}
\def\ath{Athena\xspace}
\def\mpbh{M_{\rm pbh}}
\def\fpbh{f_{\rm pbh}}

\usepackage[utf8]{inputenc}

\begin{document}

\title{
Search for primordial black hole dark matter with X-ray spectroscopic and imaging satellite experiments and prospects for future satellite missions}

\author{Denys Malyshev}
\affiliation{\!\!\mbox{ \footnotesize Institut f\"ur Astronomie und Astrophysik, Universit\"at T\"ubingen, Sand 1, D 72076 T\"ubingen, Germany
}\,\,\,\vspace{0.7ex}}

\author{Emmanuel Moulin}
\affiliation{\!\!\mbox{ \footnotesize IRFU, CEA, D\'epartement de Physique des Particules, Universit{\'{e}} Paris-Saclay, F-91191 Gif-sur-Yvette, France}\,\,\,\vspace{0.7ex}}

\author{Andrea Santangelo}
\affiliation{\!\!\mbox{ \footnotesize Institut f\"ur Astronomie und Astrophysik, Universit\"at T\"ubingen, Sand 1, D 72076 T\"ubingen, Germany
}\,\,\,\vspace{0.7ex}}

%s\author{?}

\date{\today}

\begin{abstract}
Ultra-light primordial black holes (PBHs) in the mass range of 10$^{16}$ - 10$^{22}$\,g are allowed by current observations to constitute a significant fraction, if not all, of the dark matter in the Universe.
In this work, we present limits on ultra-light, non-rotating PBHs which arise from the non-detection of the Hawking radiation signals from such objects in the keV-MeV energy band. Namely, we consider %limits implied by
observations from the current-generation missions \xmm and \spi and discuss the observational perspectives of the future missions \ath, \extp, and \ths for PBH searches. Based on $3.4$\,Msec total exposure time \xmm observations of Draco dwarf spheroidal galaxy, we conclude that PBH with masses $\lesssim 10^{16}$\,g can not make all dark matter at 95\% confidence level. Our ON-OFF-type analysis of $>100$~Msec of \spi data on the Milky Way halo puts significantly stronger constraints. Only $\lesssim 10$\% dark matter can be presented by PBHs with masses $\lesssim 3\cdot 10^{16}$\,g while the majority of dark matter can not be {represented} by PBHs lighter than $7\cdot 10^{16}$\,g at 95\% confidence level. We discuss the strong impact of systematic uncertainty related to the variations of instrumental and astrophysical \spi background on the derived results and estimate its level. 
We also show that future large-field-of-view missions such as \ths/X-GIS will be able to improve the constraints by a factor of $10-100$ depending on the level of control under the systematics of these instruments.
\end{abstract}

%\pacs{95.35.+d, 95.55.Ka, 98.56.Wm, 07.85.-m}

\maketitle

\section{Introduction}
The interest in black holes as a macroscopic dark matter (DM) candidate has been revived in last years in light of the detection of gravitational waves~\cite{LIGOScientific:2016aoc} and the strong constraints on the simplest elementary particle candidates obtained from collider searches, direct and indirect detection. Many scenarios predict the formation of primordial black holes (PBH) in the early universe and suggest that the fraction of dark matter built of PBHs, $\fpbh$ can be close to 1~\cite{Carr:2016drx,khlopov10,Green:2020jor}.

The $\mpbh$-$\fpbh$ parameter space of the PBH dark matter is strongly constrained for the small ($\mpbh\lesssim 10^{15}$\,g) and high ($\mpbh\gtrsim 10^{34}$\,g) PBH masses~\cite[see e.g.][and references therein]{villanueva21}. The PBHs with masses $\mpbh\lesssim 10^{15}$\,g would have been completely evaporated since the Big Bang by now. This opens final-stage emission searches for high-energy photon bursts expected before $\sim$10$^{15}$\,g mass PBHs completely evaporate~\cite{Linton:2006yu,Kumar:2019fqk,HAWC:2019wla,HESS:2021rto}.
The limits on the very massive PBHs with $\mpbh\gtrsim 10^{34}$\,g arise from non-observations of PBH-accretion signatures on CMB~\cite{serpico20}. Strong constraints based on non-observation of microlensing events in nearby galaxies are set in the mass range $\mpbh\gtrsim 10^{22}$\,g~\cite[see, e.g., Ref.][and references therein]{villanueva21}.

The mass range $\mpbh\sim 10^{16}-10^{21}$\,g currently remains the only relatively weakly explored window in the PBH $\mpbh-\fpbh$ parameter space. The existing constraints are concentrated at the lower part of this band ($\mpbh\sim 10^{15}-10^{18}$\,g) and are based on non-observation of the Hawking radiation's signatures from PBHs evaporation from certain astrophysical objects in the keV-MeV energy band~\cite[see, e.g., Ref.][for a review]{auffinger22}. These include constraints based on extragalactic cosmic X-ray diffuse background observations~\cite{Arbey:2019vqx,Ballesteros:2019exr,Carr:2020gox}; keV-MeV surveys of the inner parts of the MW or dwarf spheroidal galaxies~\cite{laha20, laha21, berteaud22, siegert22} ; electron-positron (511\,keV) annihilation line observations in the Galactic Center vicinity~\cite{Bambi:2008kx,Boudaud:2018hqb,DeRocco:2019fjq,laha19} and CMB power spectrum {and 21~cm signal} distortion measurements~\cite{Poulter:2019ooo,Acharya:2019xla,Acharya:2020jbv, saha22}. 
We additionally note that the strong GRB femtolensing constraints present at the lower edge of this band were recently debated and significantly relaxed (see, e.g., Ref.~\cite{Katz:2018zrn}).

Aiming {at} putting constraints on the fraction of dark matter that could be made of PBHs in this paper we present results from the current generation and discuss the potential of future missions for such studies. In what {follows} we focus on two major types of instruments characterized either by a large effective area (\xmm or \ath) or a broad field of view (FoV) -- \spi or \ths/XGIS. We discuss the most suitable observational targets and the impact of the systematic uncertainties connected to the mis-modeling of the instrumental/astrophysical background on the derived results.

The paper is organized as follows. Section~\ref{sec:signals} provides a short recap of the expected keV-MeV signals expected from Hawking radiation of PBHs in the dark matter halo of the Milky Way (MW) and  nearby dwarf spheroidal galaxies (dSphs) assuming monochromatic PBH mass function. In Section~\ref{sec:XMM_Draco},
we analyze deep observations of the Draco dSph taken by \xmm satellite. 
Section~\ref{sec:INTEGRAL_innerMW} is devoted to the analysis 
of the massive dataset taken by the high-spectroscopic-performance instrument SPI on board the INTEGRAL satellite. Prospective studies are carried out in Section~\ref{sec:prospects} to derive the sensitivity of future missions. The results presented here are discussed in Sec.~\ref{sec:discussion}.

\section{Expected keV-MeV signals from Primordial black holes}
\label{sec:signals}

\subsection{Emission spectrum}
The expected particle yield per unit time and energy from a non-rotating black hole with mass $M_{\rm BH}$ and corresponding Hawking temperature $T_{\rm H} = 1/(4\pi/G_{N\rm }M_{\rm BH})\simeq 1.06\times(10^{16} \rm g / M_{\rm BH})$ MeV, where $G_{\rm N}$ denotes the Newton's gravitational constant, is given by~\cite{Hawking:1974rv}:
\begin{equation}
\frac{d^2N_{\rm k}}{dE_{\rm k}dt} = \frac{1}{2\pi}\frac{\Gamma_{\rm k}(E_{\rm k},M_{\rm BH},m)}{e^{E_{\rm k}/T_{\rm BH}}-(-1)^{2s}}\, .  
\label{eq:flux}
\end{equation}
$\Gamma(E,M)$ is the particle-dependent grey-body factor and
$E_{\rm k}$ indicates the energy of the emitted particle $k$ of mass $m$ and spin $s$.
Assuming that PBHs have a monochromatic mass function
and trace the
DM spatial distribution, the decay of unstable particles emitted during the radiation process produces secondary stable particles including photons in the final state.
The energy-differential flux of photons expected from PBH DM halos from a region of solid angle $\Delta\Omega$ in the sky is obtained by summing all the photons produced  in the final state by all particles produced in the evaporation process as:
\begin{equation}
 \frac{d^2\Phi_{\gamma}}{dE_{\gamma}}(\Delta \Omega) = \frac{1}{4\pi} \int\limits_{\rm \Delta\Omega}d\Omega \int\limits_{\rm LOS} ds \frac{\fpbh\,\rho_{\rm DM}(r(s, d, \theta))}{\mpbh} \frac{d^2N_{\gamma}}{dE_{\gamma}dt} \, ,
\label{eq:fluxonearth}
\end{equation}
%where $\fpbh$ represents the fraction of the DM in PBHs.
The latter equation contains a factor identical to that derived in decaying DM searches and is usually referred as to the $D$-factor given by: 
\begin{equation}
  D(\Delta \Omega) = \int\limits_{\Delta \Omega} \int\limits_{\rm LOS} \: \rho_{\rm{DM}}(r(s, d, \theta)) \:ds \: d\Omega.   \label{eq:Dfactor} \\
\end{equation}
The photon emission peaks at an energy $E_\gamma\simeq 5.7\,T_{\rm H}$~\cite{MacGibbon:2007yq} and decreases as a power law for $E_\gamma \ll T_{\rm H}$.

The public software BlackHawk~\cite{Arbey:2019mbc,Arbey:2021mbl} is used to calculate the spectra of photons between 1 keV and 1 MeV. BlackHawk includes the computation of the secondary particle production due to hadronization, fragmentation, decay, and other processes as a result of BH evaporation. {In this work we make use of \texttt{BlackHawk~v2.0}~\cite{Arbey:2021mbl}. The determination of the secondary spectra depends on the evolution of Standard Model particles emitted from Hawking radiation. Public codes such as PYTHIA~\cite{Sjostrand:2014zea} and HERWIG~\cite{Bellm:2019zci} enables to convolve the primary spectra with hadronization and decay branching ratios from a few GeV up to about 10 TeV. 
For the computation of the spectrum of lower-energy photons, \textit{i.e.}, below the QCD scale, pions are emitted instead of single quarks, which subsequently decay into leptons and photons. The public 
code~\texttt{Hazma}~\cite{Coogan:2019qpu} handles the behaviour of the particles at low energy to evolve the primary particles and recover the secondary photon spectra in the keV-MeV energy range. However, \texttt{Hazma} calculations are based on analytical formulas for the decay and final-state radiations,  which introduces plausible approximations as pointed out in Ref.~\cite{Auffinger:2022dic}.}

%%%%%%%%%%%%%%%%%%%%%%%%%%%%%%%%%%%%%%
\begin{figure*}[!ht]
\includegraphics[width=0.48\linewidth]{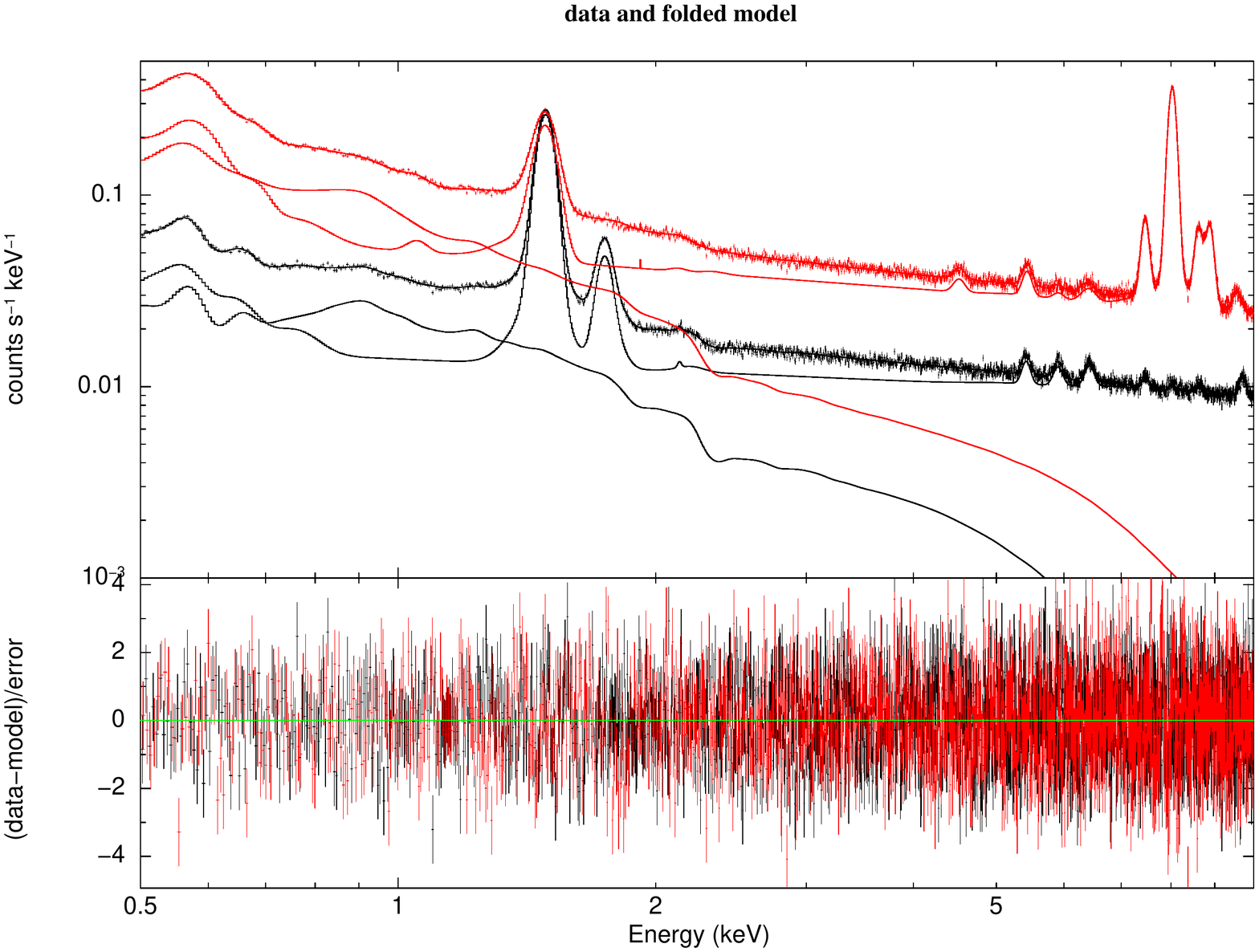}
\includegraphics[width=0.48\linewidth]{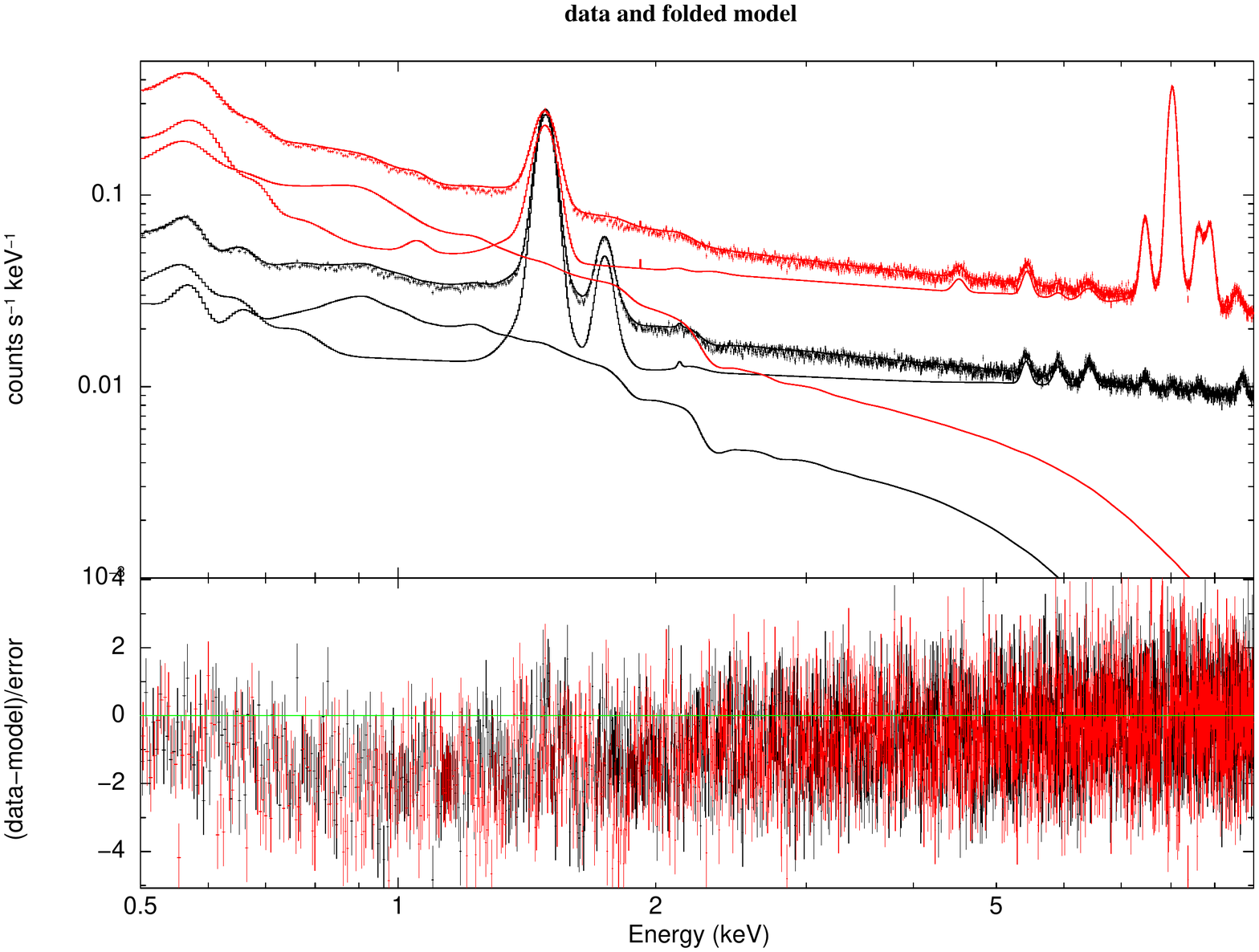}
\caption{{\it Left panel:}\xmm spectrum (top) extracted from Draco dSph region and residuals with the best-fit background model (bottom). {\it Right panel:} same with the normalization of signal from PBHs set to $2\sigma$ excluded value for a PBH mass of $\mpbh=2\cdot 10^{16}$~g. Red and black data points correspond to the stacked data of 2.5 Ms \xmm/MOS1+2 and 0.9 Ms \xmm/PN cameras, respectively.
}
\label{fig:xmm_spectrum}
\end{figure*}
%%%%%%%%%%%%%%%%%%%%%%%%%%%%%%%%%%%%%%

\subsection{Dark matter distribution}
As we discussed above the expected signal from evaporating primordial black holes is proportional to the $D$-factor in the FoV of the considered instrument. In order to optimize the signal-to-noise ratio and capabilities of the instrument for the dark matter searches a target with an angular size comparable to the FoV of the instrument is usually selected. In what {follows} we discuss the capabilities of existing and future narrow (degree-scale) FoV and broad (steradian-scale) FoV missions for the search for PBH signals. Correspondingly we select dwarf spheroidal galaxies for narrow and the Milky Way for broad FoV as the optimal targets. We also discuss the dark matter distribution in these types of objects and the uncertainties which it implies on the $D$-factor.
\paragraph{Dwarf spheroidal Galaxies.}

dSphs provide a promising  astrophysical environment to test the nature of dark matter. Due to their relative proximity and dense environments, they are amongst the best location to indirectly search for non-gravitational  interaction of dark matter.  Given their old population of stars and their low gas content, they are ideal places to look for DM where no conventional astrophysical emissions have been detected so far.

The knowledge of the DM distribution in dSphs is subject to intense studies, see, for instance, Refs.~\cite{Geringer-Sameth:2014yza,Bonnivard:2015xpq,Evans:2016xwx,Pace:2018tin}. 
Using stellar kinematic measurements from optical observations in dSphs, the DM distribution in these objects can be inferred~\cite{Geringer-Sameth:2014yza,Bonnivard:2015xpq}. 
For the nearby faintest dSphs such as Segue I, 
the selection of member stars for the faint systems may be also challenging due to the
complexity to distinguish member stars from interlopers in
the foreground. Possible tidal effects from MW and binary star population would artificially inflate the velocity dispersion and therefore the determination of their DM content (see, for instance, Ref.~\cite{Bonnivard:2015vua}).
While DM signals from the classical dSphs such as Draco, Sculptor or Fornax are expected to be smaller than for the faintest ones, they are less prone to statistical and systematic uncertainties. 
In particular, independent $D$-factor estimates for Draco dSph in a solid angle of 0.5$^\circ$ are $\rm log_{10}\,D(<0.5^\circ) = 18.53^{+0.10}_{-0.13}$\,\rm GeV cm$^{-2}$~\cite{Geringer-Sameth:2014yza}, $18.39^{+0.25}_{-0.25}\, \rm GeV cm^{-2}$~\cite{Evans:2016xwx} and $18.54^{+0.11}_{-0.14}$\, \rm GeV cm$^{-2}$~\cite{Pace:2018tin}. 
In the case of the ultra-faint dSph Segue I, independent $D$-factor estimates are 
$\rm log_{10}\,D(<0.5^\circ) = 17.99^{+0.20}_{-0.31}$\,\rm GeV cm$^{-2}$~\cite{Geringer-Sameth:2014yza} and  
$18.17^{+0.39}_{-0.39}$\,\rm GeV cm$^{-2}$~\cite{Evans:2016xwx}.

\paragraph{Milky Way Galaxy.}
DM signals from the central region of the Milky Way are expected to be stronger than that from the dSphs. However, the determination of the DM distribution in the central region of the MW
is not firmly predicted neither from  mass-modeling approaches nor from cosmological hydrodynamical simulations. Nevertheless, while the imperfect knowledge of the DM distribution in the central region of the Milky Way leads to significant uncertainty in the case of DM annihilation signal searches, its impact is significantly reduced in the case of PBH dark matter searches due to the dependence of the expected signal on the $D$-factor.
In what follows, we make use of a recent Milky Way mass model referred {to as} NFW extracted from Ref.~\cite{Cautun:2019eaf}.

For all estimates of the constraints on $\fpbh$ presented below we used the values of $D$-factors summarised in Tab.~\ref{tab:future_missions} based on dark matter profiles reported in 
Ref.~\cite{Geringer-Sameth:2014yza} (for dSphs) and Ref.~\cite{Cautun:2019eaf} for the MW and/or MW contribution to the dSphs observations with narrow-FoV instruments.

\section{\xmm constraints from Draco dwarf spheroidal galaxy}
\label{sec:XMM_Draco}
%\subsection{Observations and data set}
As described in the introduction of this paper we present the constraints on the fraction of PBH dark matter $\fpbh$ based on the analysis of current (\xmm and \spi) and future (\extp, THESEUS, Athena) missions. 

\xmm is currently operational, state-of-the-art X-ray mission, operating at energies $\sim 0.2 - 12$~keV and equipped with MOS and PN cameras characterised by a high total effective area peaking at $\sim 2000$\,cm$^2$ at 1.5~keV, a good energy resolution of $\sim 10$\% and relatively broad FoV of $\sim 15'$ radius~\cite{xmm_pn}. In terms of the proposal devoted to deep studies of the decaying sterile neutrino in 2015 \xmm performed a deep dedicated observation of Draco dwarf spheroidal galaxy. The obtained dataset was extensively used in numerous works devoted to indirect sterile neutrino searches (see, for instance, Refs.~\cite{Ruchayskiy:2015onc,Jeltema:2015mee,Ruchayskiy:2015onc}). 

In this paper, we present the re-analysis of these data following the data-reduction scheme used in Ref.~\cite{Ruchayskiy:2015onc}. Namely, we analyzed the data with the Extended Source Analysis Software (ESAS) included in XMM-SAS software v.19.1.0 with the most recent calibration files. The time intervals strongly affected by soft proton flares were removed with the help \texttt{mos-filter} ESAS script with the standard cuts. For additional minimization of the astrophysical background, we masked out the point-like sources in the FoV with \texttt{cheese} ESAS procedure. The selected cut-out radius ($36''$, similar to~\cite{Ruchayskiy:2015onc}) allows us to remove up to $70$\% of the point source flux. The spectra and response matrices for individual observations were produced by \texttt{mos-spectra} ESAS routine. The spectra of individual observations for MOS and PN cameras were stacked together with the help of \texttt{addspec} and binned to 65~eV bins with \texttt{grppha} FTOOL. The total clean exposures of MOS and PN spectra are 2.5~Msec and 0.9~Msec, respectively.

The resulting spectra consist dominantly of the astrophysical (solar system plasma, hot interstellar plasma,  cosmic X-ray background) and the instrumental (smooth continuum and line-like features) backgrounds. Following Ref.~\cite{stacked_dsph_sterile} we modelled the astrophysical background with \texttt{apec + TBabs(apec+powerlaw) } XSpec model presenting contributions from the described components. The instrumental background was modeled as a (not convolved with the effective area)  power-law spectrum. We additionally included into the instrumental background model Gaussian lines, 
 reported in Ref.~\cite{leccardi08} (Table A.2), Ref.~\cite{snowden04} (Table 2) and $4.512$~keV keV line to model Ti~K$\alpha$ instrumental line~\cite{bulbul20}, see Tab.~\ref{tab:xray_lines} for the list of all lines included into the model. The suggested signal from the evaporating PBHs was added as an additive table model component with a free normalization to the astrophysical background model.

\begin{table}
    \centering
    \begin{tabular}{c|c|c}
        \hline         \hline  
        Energy&Line&Origin\\
        \hline 
        %0.37& C VI& astrophysical line\\
        %0.46& C VI& astrophysical line\\
        0.56&O VII& astrophysical line\\
        0.65&O VIII& astrophysical line\\
        0.81&O VIII& astrophysical line\\
        0.91&Ne IX& astrophysical line\\
        1.34&Mg XI  & astrophysical line\\      
        1.49& 	Al-K$\alpha$& instrumental line\\
        1.56&    Al-K$\beta$&  instrumental line\\
        1.74&    Si-K$\alpha$&  instrumental line\\
        1.84& 	Si-K$\beta$& instrumental line\\
        2.11&    Au-M$\alpha$ &  instrumental line\\
        2.20&    Au-M$\beta$ &   instrumental line\\
        4.51& 	Ti-K$\alpha$& instrumental line\\
        5.41& 	Cr-K$\alpha$& instrumental line\\
        5.89&    Mn-K$\alpha$&  instrumental line\\
        5.95&    Cr-K$\beta$ &  instrumental line\\
        6.40& 	Fe-K$\alpha$& instrumental line\\
        6.49&    Mn-K$\beta$  &  instrumental line\\
        7.06&    Fe-K$\beta$  &  instrumental line\\
        7.48& 	Ni-K$\alpha$& instrumental line\\
        8.04& 	Cu-K$\alpha$& instrumental line\\
        8.26&  	Ni-K$\beta$  &  instrumental line\\
        8.63& 	Zn-K$\alpha$& instrumental line\\
        8.90& 	Cu-K$\beta$& instrumental line\\
        9.57& 	Zn-K$\beta$& instrumental line\\
        9.68&    Au-L$\alpha$&  instrumental line\\
                \hline         \hline 
    \end{tabular}
    \caption{The most prominent instrumental and astrophysical lines used for the \xmm background modelling. The lines are adopted from Ref.~\cite{leccardi08} (Table A.2), Ref.~\cite{snowden04} (Table 2) and $4.512$~keV line is from~\cite{bulbul20}.}
    \label{tab:xray_lines}
\end{table}

\begin{figure}
\includegraphics[width=\linewidth]{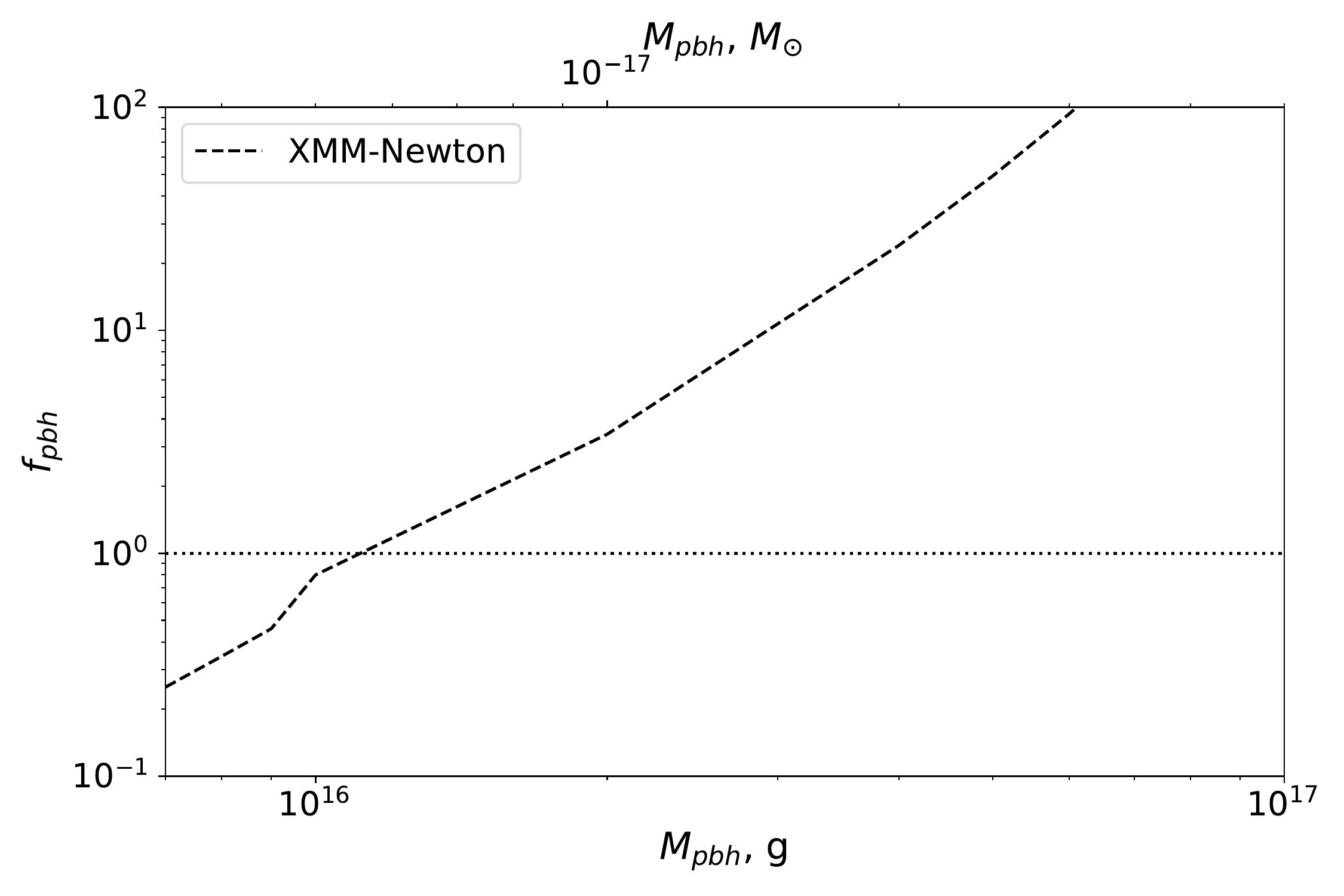}
\caption{95\% C.L.  upper limits on the fraction of PBHs $\fpbh$ based on stacked 2.5~Msec \xmm/MOS(1+2) and 0.9~Msec \xmm/PN of Draco dSph observations.
}
\label{fig:xmm_limits}
\end{figure}

The described model provides a good fit to the data (c-statistic is used during the fit\footnote{\href{https://heasarc.gsfc.nasa.gov/xanadu/xspec/manual/XSappendixStatistics.html}{See description of statistics used in XSpec.}}), as shown in Fig.~\ref{fig:xmm_spectrum}. It allows us to put constraints on the normalisation of the searched signal from evaporating PBHs and consequently to $\fpbh$. The presented limits are computed with \texttt{error 4.0} Xspec command and correspond to $2\sigma$ confidence level (C. L.) upper limits. The limits on $\fpbh$ as a function of PBH mass are shown in Fig.~\ref{fig:xmm_limits}.

\section{Constraints from \spi observations of the Milky Way}
\label{sec:INTEGRAL_innerMW}
%\subsection{Observations and data set}
\spi is a coded-mask instrument on board of INTEGRAL satellite~\cite{winkler03,2kuulkers21} operating in $\sim 0.1-10$~MeV energy band. The instrument is characterised by a relatively high effective area ($\gtrsim 100$~cm$^2$ at 100~keV) and excellent energy resolution ($\Delta E/E\sim 1/500$)~\cite{spi_description}. The satellite is located on $\sim 3$-day period, highly eccentric orbit which is partially located in the Earth's radiation belts. The regular crossage of the belts leads to a strong irradiation of the satellite by high-energy charged particles and consecutively to a strong time-variable instrumental background. 

Contrary to \xmm, the absence of focusing optics in \spi greatly reduces the imaging capabilities of the instrument. For the robust analysis presented below, we explicitly use \spi as a collimator with a $17.5^\circ$-radius (partially-coded) FoV, similar to the analysis performed in Ref.~\cite{we_spi}. Contrary to a template-based analysis of Ref.~\cite{berteaud22}, our analysis results in a somewhat increased background level (connected to, \textit{e.g.}, contributions from the point-like sources in the FoV). On the opposite, %such an
our analysis allows us to firmly estimate the level of systematic uncertainties connected to the time-variable instrumental background.

%%%%%%%%%%%%%%%%%%%%%%%%%%%%%%%%%%%%%%
%%%%%%%%%%%%%%%%%%%%%%%%%%%%%%%%%%%%%% 

We selected for the analysis all publicly available\footnote{\spi observations are available from \href{https://www.isdc.unige.ch/integral/archive}{ISDC website.}} pointing \spi observations performed between Nov. 4th, 2002 and Oct. 16th, 2021 (ScWs: 000752000100 -- 242300650010) taken in normal operating mode (\texttt{spimode=41} data selection flag) with, at least, 100~s {good\_spi} exposure. 

Aiming at PBH DM searches in the MW characterized by dark matter density profile decaying with the distance from the GC,
we split the available observations over ``ON'' and ``OFF'' groups. We select ON-observations to be performed within $60^\circ$ away from the GC, while for OFF observations -- those to be performed at $>90^\circ$ offset from the Galactic Center. Such a definition of ON and OFF regions allows us to expect higher signals from evaporating PBHs in ON region and use the OFF region to control the time-variable instrumental background. To minimize the effects of the time-dependent background variability we additionally divided ON and OFF observations into 
sub-groups close in time. Namely, we split all ON observations over groups of a maximal duration of 20~days. Similarly, we split all OFF observations over groups of $\leq 20$~days with an additional requirement that any observation in OFF-group shall not be taken later than 20~days after the first observation in the corresponding ON-group. 

Such additional sub-division and consequent sub-selection allow us for accurate control of the time-variable \spi instrumental background. As shown in Ref.~\cite{tw06} for observations separated in time by not more than 20~days, the level of instrumental background is in a good correlation with a strength of a strong instrumental Germanium line at 198~keV. In this case the strength of the line can be used for the renormalization of the \spi instrumental background. Consequently, the expected signal from the evaporating PBHs in our analysis is given by
\begin{align}
& \frac{d^2\Phi_\gamma}{dE_\gamma dt} = \frac{\fpbh}{4\pi \mpbh}\frac{d^2N_\gamma}{dE_\gamma dt}\sum\limits_{\rm i}\left( D_{\rm ON,i} - \alpha_iD_{\rm OFF,i}\right) \\ \nonumber
& {\rm with\quad} \alpha_{\rm i} = f_{\rm ON,i}(198\,\mbox{keV}) / f_{\rm OFF,i}(198\,\mbox{keV})\, .
\label{eq:spi_signal}
\end{align}
The index $i$ corresponds to sub-groups of ON/OFF observations, $f_{\rm ON, i}(198\,\mbox{keV})$ and  $f_{\rm OFF, i}(198\,\mbox{keV})$ to the flux in the 198~keV instrumental line in the ON  and OFF regions of the $i$-th %ON and OFF 
group, respectively. We note that all obtained $\alpha_i$ during the analysis deviate from 1 by 10\% at most, typically within a few percent. 
With the setup of the analysis described above, we were able to identify 216 %groups of 
ON-OFF pairs with the total ON and OFF exposures of $126$~Ms and $94$~Ms, respectively. The exposure-averaged $D$-factors of the ON and OFF regions are $1.1\cdot 10^{22}$~GeV/cm$^2$ and $0.2\cdot 10^{22}$~GeV/cm$^2$, respectively, for the considered NFW profile from the MW halo of Ref.~\cite{Cautun:2019eaf}.

%%%%%%%%%%%%%%%%%%%%%%%%%%%%%%%%%%%%%%%%%%%%%%%%%%%%%%%%%%%
\begin{figure}
    \centering
    \includegraphics[width=\linewidth]{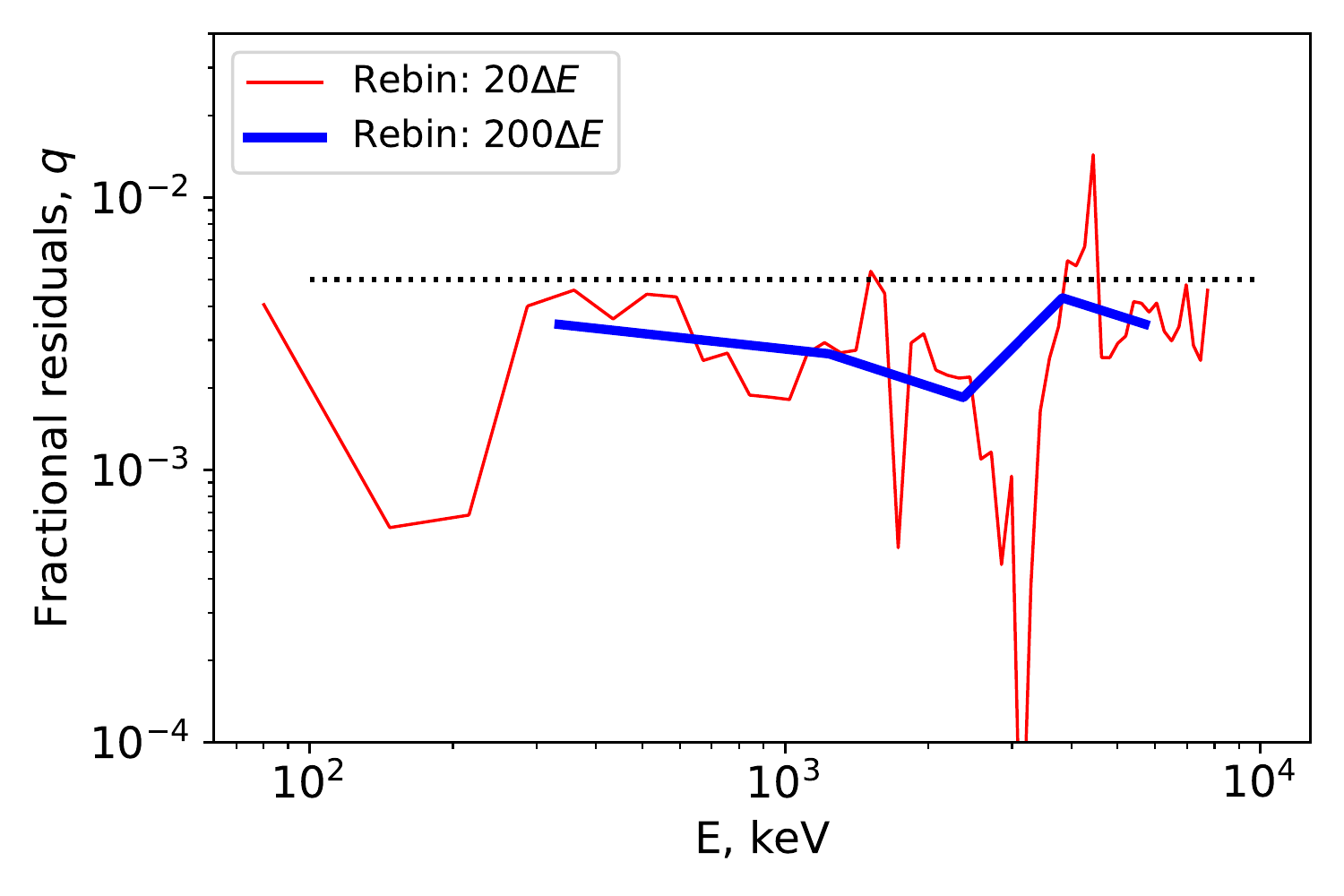}
    \caption{The systematic uncertainty of the ``ON-OFF'' analysis of \spi data presented in this work from 126 Ms and 94 Ms of MW observations by \spi for the ON and OFF regions, respectively. The curves show the fractional residuals ($q\propto (N_{ON}-N_{OFF})/N_{ON}$) for the data binned in narrow energy bins ($20\Delta E$ width, red curve) and broad energy bins ($200\Delta E$, blue curve).}
    \label{fig:spi_systematics}
\end{figure}
%%%%%%%%%%%%%%%%%%%%%%%%%%%%%%%%%%%%%%%%%%%%%%%%%%%%%%%%%%%

In order to extract the spectra of individual groups of observations we extracted the detected photons' list from the \spi event files (\texttt{spi\_oper*.fits.gz}) and initially binned the obtained list to narrow, $\sim 1/5$ of \spi energy resolution, energy bins. The \spi energy resolution is given by~\cite{we_spi}:
\begin{align}
& \frac{\Delta E(E)}{1\,\mbox{keV}} = 1.54 + 4.6\cdot 10^{-3}\sqrt{\frac{E}{\mbox{1\,keV}}} + 6\cdot 10^{-4} \left(\frac{E}{\mbox{1\,keV}}\right) \, .
\end{align}
Such selection of bins allowed us to accurately estimate the flux in the narrow instrumental 198~keV line, determine $\alpha_i$ and perform proper spectra rescaling of OFF groups of observations. 

The discussed procedure allowed us {to make} an accurate subtraction of the spectra of ON and OFF regions. We estimated the accuracy of subtraction by a spectral fractional residual $q = abs(\sum (ON_i - \alpha_i OFF_i) )/ \sum ON_i$. This quantity as a function of energy is shown in Fig.~\ref{fig:spi_systematics} for differently re-binned ON and OFF spectra. The rapidly oscillating red curve illustrates that close to the complex regions dominated by a blend of strong instrumental lines the adapted background subtraction procedure operates worse than in the continuum. Particularly strong variations could be seen close to $E\sim 1434-1460$~keV ($^{52}V^{52}Cr$ + $^{40}K^{40}Ar$ blend) and $E\sim 4122-4434$~keV ($^{66}Ga^{66}Zn$ + $^{12}C$ blend)~\cite{spi_lines}.
Consideration of broader energy bins results in the smoothing of oscillations. The discussed fractional residuals allow us to estimate the level of systematic uncertainty in our analysis connected to the imperfect modeling (subtraction) of the background assuming that there is no significant excess between the ON and OFF regions. Following Fig.~\ref{fig:spi_systematics} we conservatively estimated the systematic uncertainty to be 0.5\% (of the ON region flux). In what follows we add this systematic uncertainty to the statistical one. We note also that the discussed systematic uncertainty can be either uncorrelated within nearby energy bins or, on the opposite, strongly {correlated} over the whole considered energy range. 

In the case of uncorrelated systematic uncertainty, it can be taken into account by adding in quadratures, similarly to the statistical one. The relative error on the strength of the signal present in several or many energy bins reduces with the number of bins and can be infinitely small. The strongly correlated systematic (which, \textit{e.g.}, shifts the whole ON-OFF spectrum up or down in terms of flux) on contrary should be added linearly. In this case, the relative uncertainty of the strength of the broad-band signal is limited by the level of systematic uncertainty and can not be smaller than a certain value.
Since the properties (correlated versus uncorrelated) of \spi systematic uncertainties are \textit{a priori} unknown, in what follows we consider both cases. 

\begin{figure*}
    \centering
    \includegraphics[width=0.48\linewidth]{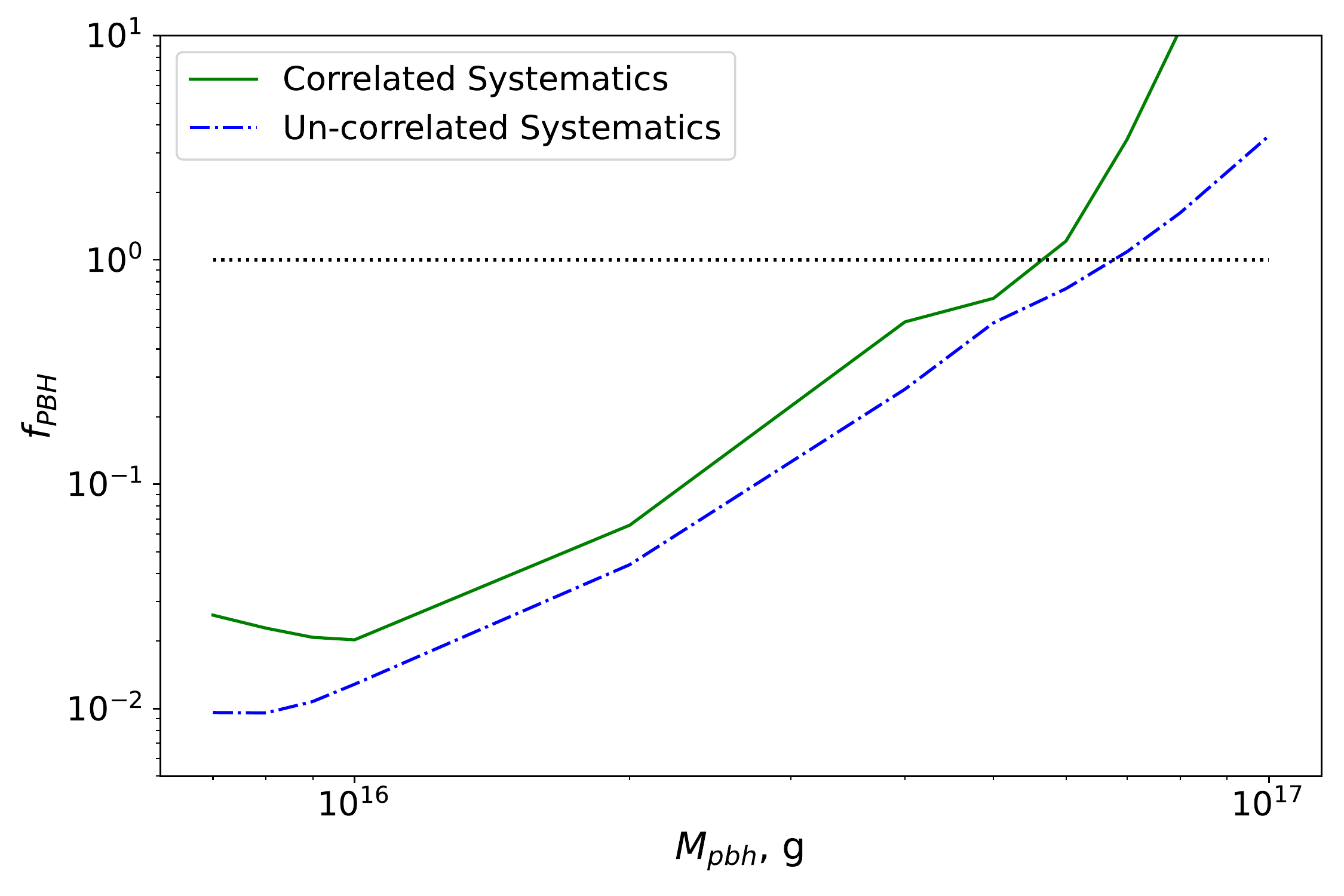}
    \includegraphics[width=0.48\linewidth]{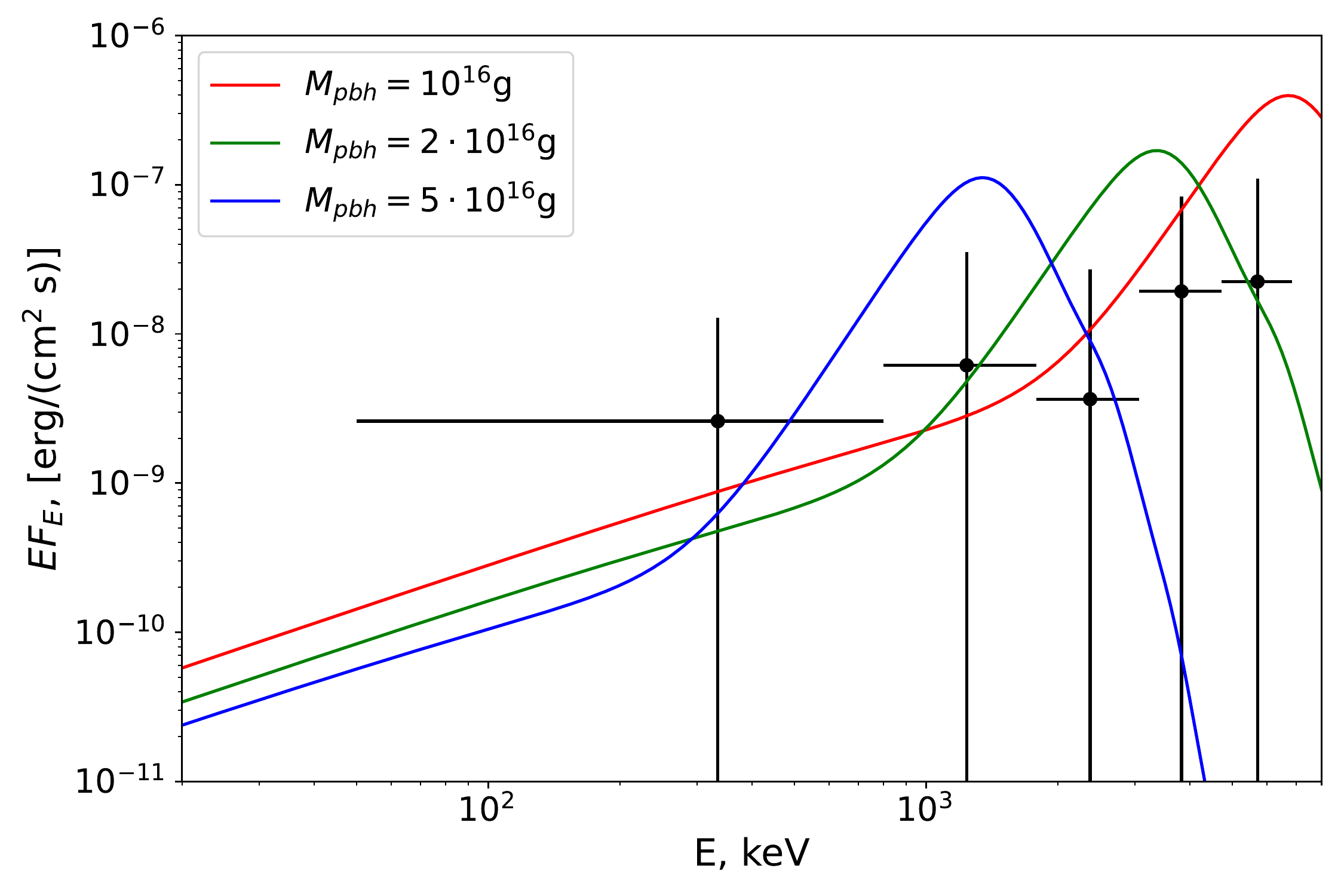}
    \caption{\textit{Left panel:} Constraints on the fraction of PBH DM, $\fpbh$, from INTEGRAL-SPI observations of the inner Milky Way. The constraints are expressed as 95\% C.L. upper limits assuming that DM can be constituted of a monochromatic PBH distribution with mass $\mpbh$. \textit{Right panel:} Examples of spectra for evaporating PBHs for PBH masses of 10$^{16}$ (red line), 2 $\cdot$ 10$^{16}$ (green line) and 5 $\cdot$ 10$^{16}$~g (blue line), respectively, and $\fpbh$ corresponding to $2\sigma$ excluded value. Black points present \spi residual spectrum binned in broad energy bins and added 0.5\% systematic uncertainty corresponding to the correlated systematic case. See text for more details. 
    }
    \label{fig:spi_constraints}
\end{figure*}

For the case of uncorrelated systematic, we added the above-mentioned 0.5\% of the total ON-region flux to the uncertainties of the residual signal. The presented 95\% C.L. limits on $\fpbh$ correspond to the signal normalization excluded by the data at  95\% C.L.
To mimic the effects of the strongly correlated systematic uncertainty we re-bin the data into broad ($\sim 200\Delta E$ width) energy bins before adding the systematics to the data. Such a choice resulted in 5 energy bins within the \spi energy range with characteristic widths of $\sim 0.3-1.5$~MeV and correspond to the correlated systematic uncertainty operating at such energy scales. After the addition of 0.5\% of the total ON-region flux to the data, similarly to the case of uncorrelated systematics, we used a $\chi^2$ test to exclude the signal of evaporating PBHs at %$2\sigma$  
95\% C.L. %significance level. 
The exclusion regions for the case of correlated systematics (blue dotted-dashed line) and uncorrelated systematics (green solid line) are shown in Fig.~\ref{fig:spi_constraints} for the difference of $D$-factors of $0.9\cdot 10^{22}$~GeV/cm$^2$ between the ON and OFF regions.

\section{Prospects for next-generation missions}
\label{sec:prospects}
In addition to the constraints on $f_{\rm{PBH}}$ delivered by currently operating \xmm and \spi satellites we extend 
our study with the sensitivity prospects for several next-generation missions which can see the first light in the next one or two decades. Namely, we consider three broadly discussed future missions such as \ath, \extp, and \ths. The instruments planned to be on board these {satellites} can be contingently divided by \xmm-like (\textit{i.e.} characterized by a relatively small FoV and a large effective area) and \spi-like (\textit{i.e.} characterized by a broad FoV and relatively small effective area) classes. Brief information on the considered missions and instruments onboard is summarized in Tab.~\ref{tab:future_missions} and discussed below.
\begin{table*}[]
    \centering
    \begin{tabular}{c|c|c|c|c|c|c|c}
    \hline\hline
    Instrument    & Energy range  & Peak $A_{\rm eff}$  & FoV & Launch date & Target & Obs. Type &$D$-factor\\
                  & [keV]           & [cm$^{2}$]        & [sr]  & [year]   &    &    & [GeV/cm$^2$]\\
                  \hline\hline
       \xmm/PN       &     0.1-15    & 815             & $4.5\cdot 10^{-5}$ & 1999-**& Draco+MW& Model  & $(1.1+0.74) \cdot 10^{18}$\\%$1\cdot 10^{18}$\\
    \spi          &     20-8000   & 160             & 0.29               & 2002-**& MW & ON-OFF     &$0.9\cdot 10^{22}$\\
    \hline
    \extp/SFA     &   0.5-10      & 8600            & $9.6\cdot 10^{-6}$ & 2027& Segue I + MW&Model& $(2.0+0.9) \cdot   10^{17}$\\%$2.84\cdot 10^{17}$ \\            
    \extp/LAD     &   2-30        & $3.3\cdot 10^4$ & $2.4\cdot 10^{-4}$ & 2027 &  Segue I& ON-OFF & $9.8\cdot 10^{17}$\\
    \extp/WFM     &     2-50      & 77              & 2.5                & 2027 &  MW & ON-OFF    & $2\cdot 10^{22}$ \\
    \ths/SXI      &     0.3-5     & 1.9             & 1                  & 2037& MW & ON-OFF     &$1\cdot 10^{22}$\\
    \ths/XGIS-X   &     2-30      & 504             & 1                  & 2037& MW & ON-OFF     &$1\cdot 10^{22}$\\
    \ths/XGIS-S   &     20-2000   & 1060            & 1                  & 2037& MW & ON-OFF     &$1\cdot 10^{22}$\\
    \ath/X-IFU    &     0.2-12    & $1.6\cdot 10^4$ & $3.3\cdot 10^{-6}$ & 2035& Segue I+MW &Model & $(8.3+3.0)\cdot 10^{16}$\\
    \ath/WFI      &     0.2-15    & 7930            & $1.35\cdot 10^{-4}$& 2035& Segue I+MW &Model &$(0.98+1.2)\cdot 10^{18}$\\%3.7\cdot 10^{18}$\\
    \hline
    \hline 
    \end{tabular}
    \caption{Technical characteristics of present-day and future missions considered in this work. The table summarizes the operating energy range of the instrument, peak effective area, the field of view, and the planned launch date (as of August 2022). The last columns summarize the target, the type of (proposed) observation (background Model or ``ON-OFF''), and the estimated $D$-factor. For the ``ON-OFF''-type observations the $D$-factor corresponds to the $D$-factor difference between the ON ad OFF regions. Adopted D-factor values are based on~\cite{Geringer-Sameth:2014yza} and~\cite{Cautun:2019eaf}.
    }
    \label{tab:future_missions}
\end{table*}

\paragraph{\extp.} The enhanced X-ray Timing and Polarimetry mission (\extp~\cite{extp_description, extp2, extp1} ) is a forthcoming\footnote{As of 2022 the launch is planned in 2027.} Chinese-European mission primarily designed for the study of the equation of state of matter within neutron stars, measurements of QED effects in highly magnetized stars, and studies of accretion in the strong-field gravity regime.

The mission will host several state-of-the-art scientific instruments operating in the soft to hard X-ray band ($0.5-50$~keV). The main instruments on board the \extp are:\\
-- The Spectroscopic Focusing Array (SFA), consisting of nine X-ray modules operating in the $0.5-10$~keV band with a field of view (FoV) of $12'$ (full-width half-maximum, FWHM), the total effective area of $\sim 0.8$~m$^2$ at 2~keV and an energy resolution of better than 10\%;\\
-- The Large Area Detector (LAD) -- non-imaging instrument operating at $2-30$~keV energies, with an FoV of $60'$ (FWHM), an effective area of $\sim 3.4$~m$^2$ and an energy resolution better than 250~eV;\\
-- The Wide Field Monitor (WFM) -- a wide, steradian-scale, FoV instrument operating in the $2-50$~keV energy band with an effective area of $\sim 80$~cm$^2$ and an energy resolution similar to that of the LAD. The capabilities of this instrument for indirect dark matter searches were recently discussed in Ref.~\cite{zhong20}.

In addition to the instruments described above, \extp will 
host the Polarimetry Focusing Array (PFA). This instrument is characterized by an effective area and a field of view similar to \xmm. Thus, in what follows we consider only SFA, LAD, and the WFM perspectives for the searches for evaporating PBHs.

We base the analysis presented below  on the simulated instrumental backgrounds and associated response files/functions given by \texttt{XTP\_sfa\_v6.bkg}\footnote{Note that the provided template corresponds to the background in $\sim 3'$ and has to be re-scaled by a factor of 16 to match $12'$ FoV of SFA. The WFM background template was provided for one module and had to be up-scaled by a factor of 3.}, \texttt{LAD\_40mod\_300eV.bkg} and \texttt{WFM\_M4\_full.bkg} templates for SFA, LAD and WFM respectively, which were provided by the \extp collaboration\footnote{See \href{https://www.isdc.unige.ch/extp/}{\extp website.}}.

\paragraph{\ths} is a European mission concept\footnote{Phase-II proposal in response to the ESA ``M7'' call is submitted in 2022.} designed in response to the ESA call for a medium-size mission (M5) within the Cosmic Vision Program\footnote{\href{https://www.esa.int/Science_Exploration/Space_Science/ESA\_s\_Cosmic\_Vision}{See Cosmic Vision Program website.}}. The fundamental goals of the \ths mission are the study and detection of high energy transient phenomena, the study of the early universe and the epoch of re-ionization, and ``the hot and energetic universe". These goals are planned to be achieved using the mission's unique combination of instruments.

 The \ths mission will host a total of three telescope arrays, covering a section of the infrared regime as well as the energy range of soft and hard X-rays. The proposed instrumental payload for \ths is:\\
 --The Soft X-Ray Imager (SXI), an array of 4 lobster-eye~\citep{angel97} telescope units with a quasi-square FoV covering the energy range of $0.3 - 5$~keV with an effective area of 
 $A_{\rm eff}
 \approx 1.9$~cm$^2$ at 1 keV and an energy resolution $\sim 4$\%. These will cover a total FoV of $\sim 1$~sr with source location accuracy $<1-2$ arcminutes (for a full review of the instrument, see Ref.~\cite{SXI}).\\
 --The InfraRed Telescope (IRT), is a single large (0.7~m) telescope that will be used for follow-up observations of gamma-ray bursts. It will operate in the wavelength band  $0.7 - 1.8\, \mu$m and have a $15'\times 15'$ FoV (for further specifications on the IRT see \cite{IRT}).\\
 --The X-Gamma Ray Imaging Spectrometer (XGIS) array, consists of coded-mask cameras (with the total half-sensitive FoV comparable to that of the SXI)  using monolithic X-gamma ray detectors based on bars of silicon diodes coupled with CsI crystal scintillator. XGIS will operate in the energy range of 2~keV -- 20~MeV, which will be achieved using the two different detectors, referenced hereafter as XGIS-X and XGIS-S. The Silicon Drift Detector (SDD) will cover the energy range of 2--30~keV (XGIS-X) whereas the CsI scintillator will cover the range of 20~keV -- 2~MeV (XGIS-S\footnote{Note, that due to the transparency of the XGIS coded mask at hard X-rays at $E\gtrsim 150$~keV XGIS-X operates as a collimator.}). The effective areas and energy resolutions of XGIS-S are $A_{\rm eff}(300\,\mbox{keV})\approx 1100$~cm$^2$ and energy resolution changing from $\Delta E/E\sim 15$\% at below $100$~keV to $\Delta E/E\sim 2$\% at higher energies. The effective area and resolution of XGIS-X instrument are $A_{\rm eff}(10\,\mbox{keV})\approx 500$~cm$^2$ and $\Delta E/E\sim 1.5$\%, see \cite{XGIS} for the full technical proposal for the XGIS.
 
 The described  simulation and analysis of \ths data below is based on templates of blank sky observations provided by the \ths collaboration\footnote{V7 templates dated May-July 2020; see \href{https://www.isdc.unige.ch/theseus/}{\ths web page.}} (\texttt{sxi\_bkg.pha}\footnote{Scaled by 17508, to account for template's FoV (675~arcmin$^2$).}, \texttt{XGIS-X\_0deg\_v7.bkg} and \texttt{XGIS-S\_0deg\_v7.bkg} ) and corresponding response files.
 
\paragraph{\ath\footnote{https://www.the-athena-x-ray-observatory.eu/}} is planned to be the second large mission (L2) launched in the framework of the Cosmic Vision program of the European Space Agency. It will host onboard X-ray telescopes with an effective area of the order of one square meter. A set of detectors in the focal plane will include a Wide Field Imager (WFI) and an X-ray Integral Field Unit (X-IFU). WFI is characterised by a relatively large FoV of $40'\times40'$, peak effective area $\sim 8000$~cm$^2$ and energy resolution of $\sim$3\%. X-IFU will have a narrower field of view of $7'$-diameter, but a significantly higher effective area (up to $16000$~cm$^2$) and spectral resolution up to $E/\Delta E \approx 2800$.

In what {follows} we used v.20210329, \texttt{*extended\_wo\_filter\_FovAvg.pha} and \texttt{Total\_1arcmin2\_XIFU\_CC\_BASELINECONF\_2018\_10\_10.pha} simulated backgrounds for the WFI and X-IFU instruments\footnote{Publicly available via: \href{http://x-ifu-resources.irap.omp.eu/PUBLIC/RESPONSES/CC_CONFIGURATION/}{X-IFU} and \href{https://www.mpe.mpg.de/ATHENA-WFI/public/resources/background/}{WFI.}}.
\begin{figure*}
    \centering
    \includegraphics[width=0.48\linewidth]{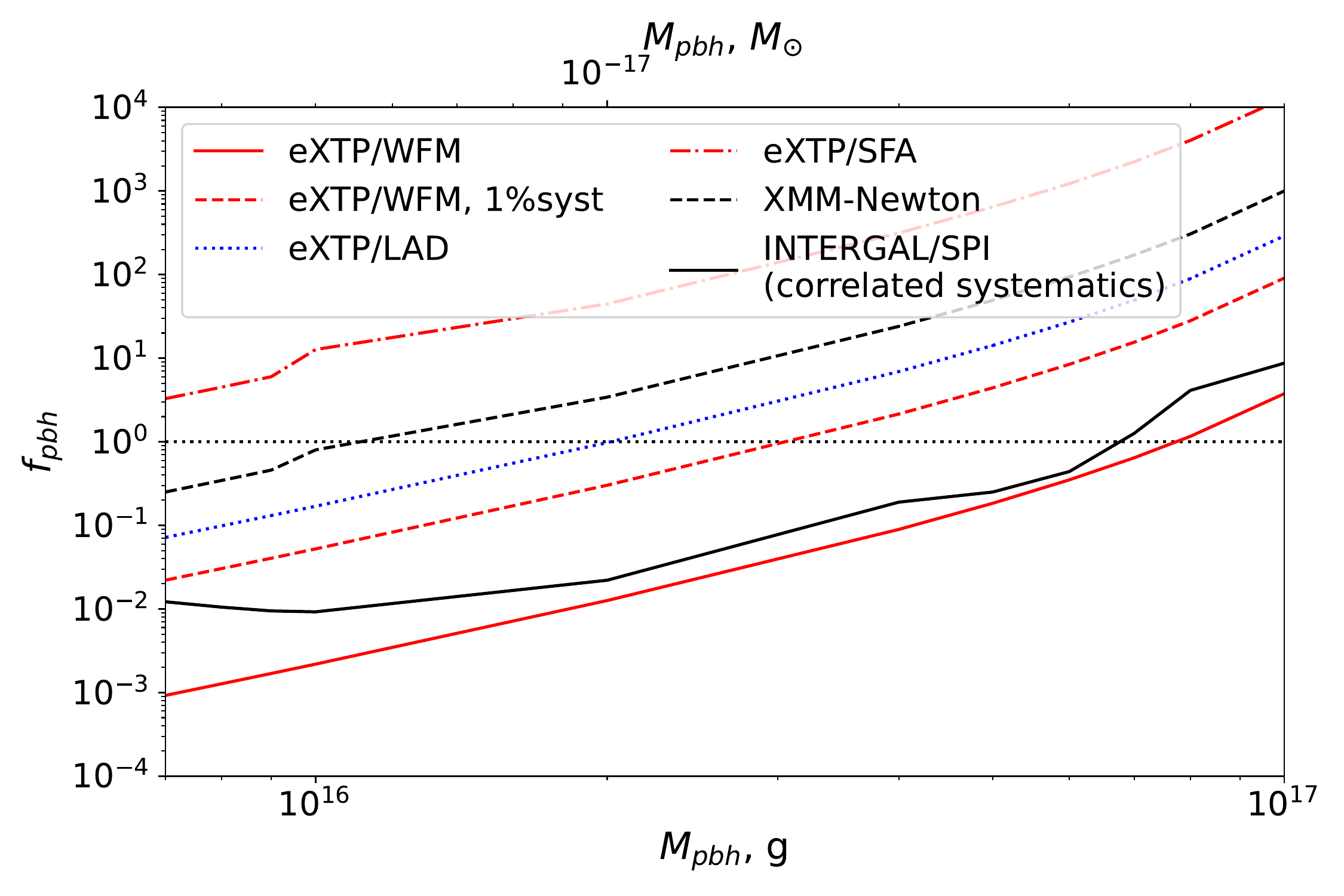}
    \includegraphics[width=0.48\linewidth]{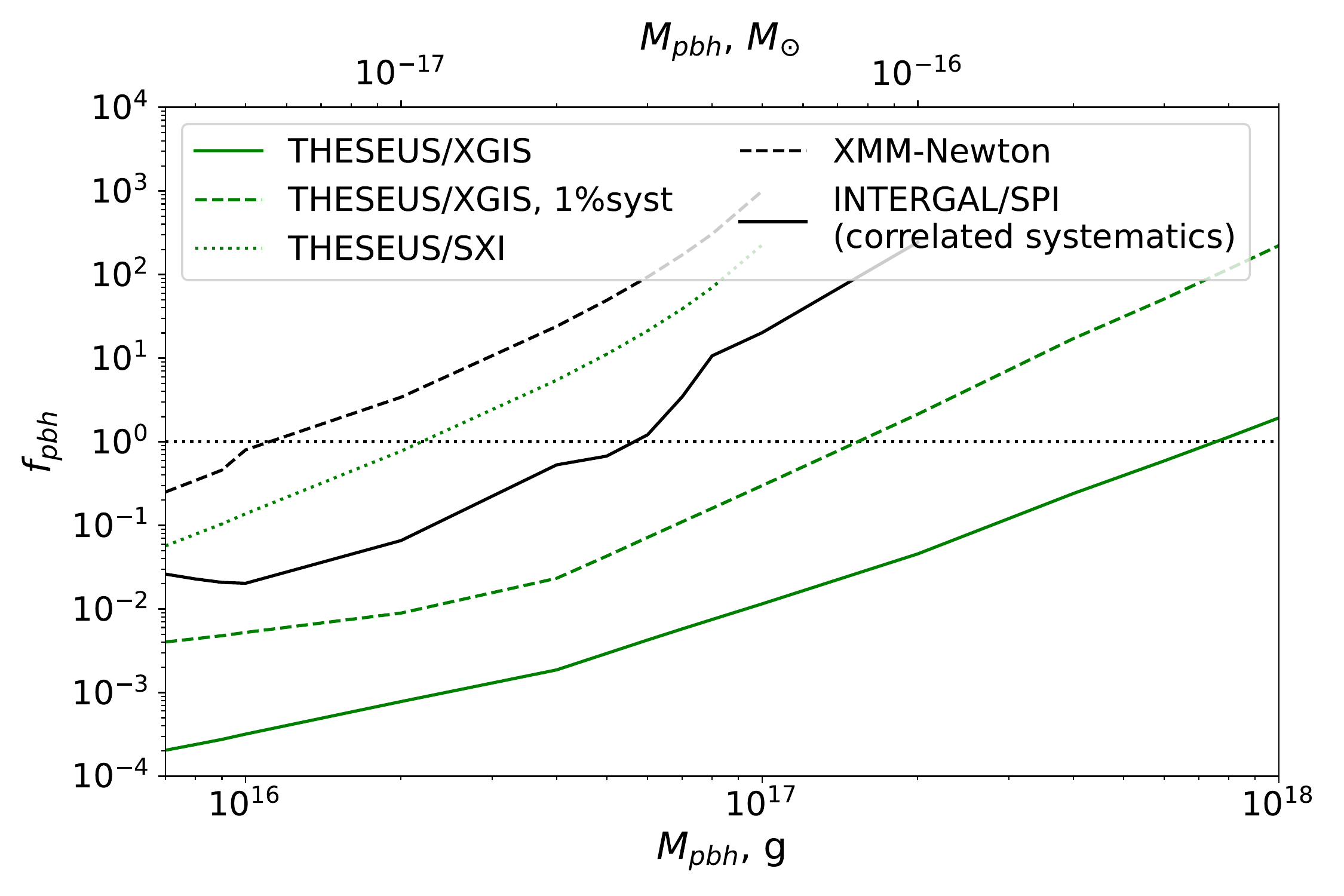}
    \caption{Constraints on the fraction of PBH DM $f_{\rm DM}$ as a function of the PBH mass $\mpbh$ from the future satellite missions  \extp and \ths (right panel; {please note different x-axis range}) in  comparison with present constraints obtained towards Draco dSph observations by \xmm and the MW galaxy with \spi. \textit{Left panel:}
    For \extp, the sensitivity is computed at 95\% C. L. for the WFM with (red dashed line) and without (red solid line) systematic uncertainty, and the LAD (blue dotted line) for 1~Msec observations of Segue I dSph. \textit{Right panel:} For THESEUS, the sensitivity is computed at 95\% C. L. for the XGIS with (green dashed line) and without (green solid line) systematic uncertainty, and the SXI (green dotted line) for 1~Msec observations of the MW galaxy. 
    }
    \label{fig:future_constraints1}
\end{figure*}

\begin{figure}
    \centering
    \includegraphics[width=\linewidth]{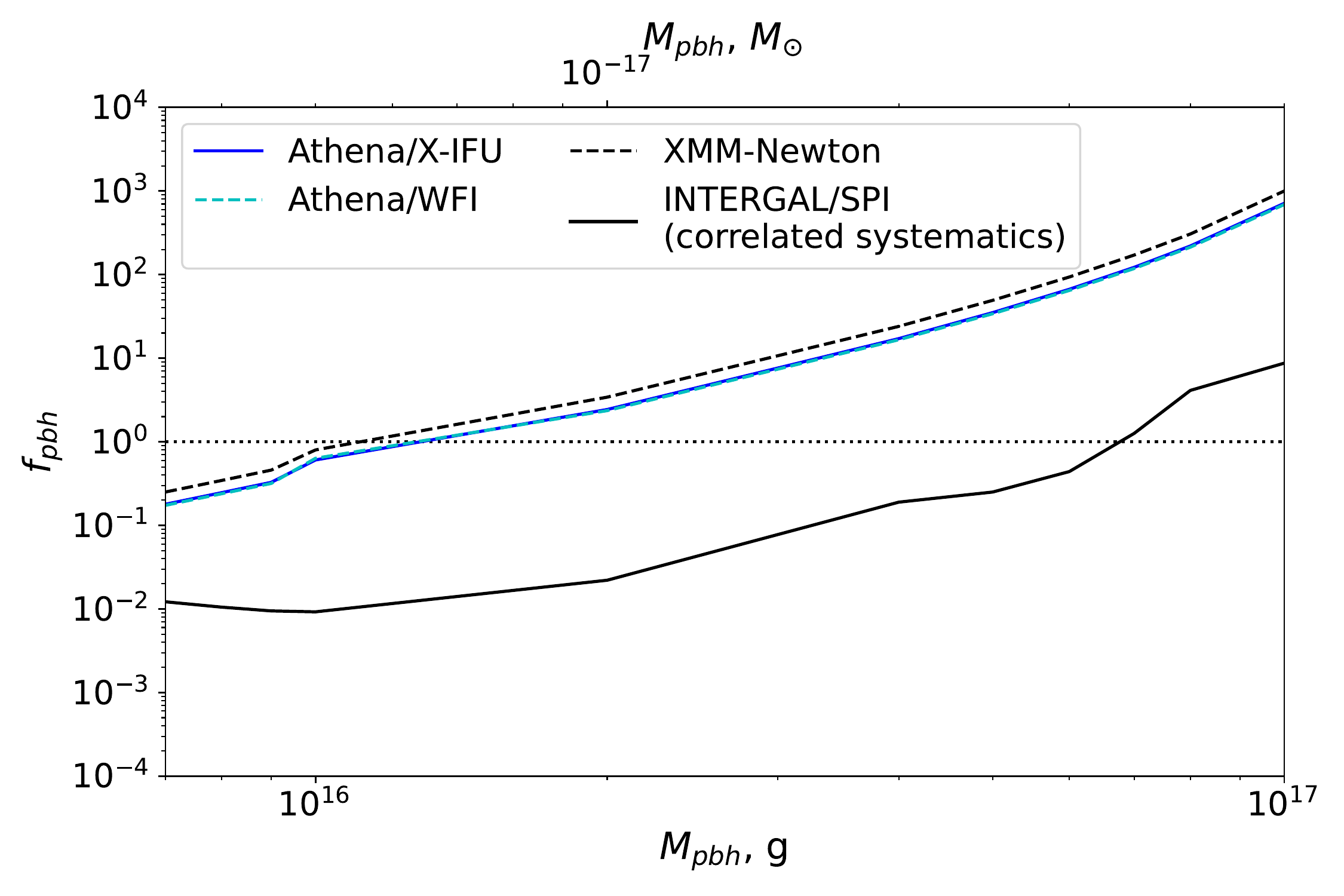}
    \caption{Constraints on the fraction of PBH DM $f_{\rm DM}$ from \ath, expressed s 95\% C.L. sensitivity, for 1~Msec of observations towards Segue I dSph with X-IFU (solid blue line) and WFI (dashed blue line) instruments in comparison with present constraints obtained towards Draco dSph observations by \xmm and the MW galaxy with \spi.}
    \label{fig:future_constraints2}
\end{figure}

Following the strategy proposed in~\cite{we_athena,we_extp,we_theseus} we propose dSphs as primary targets for narrow ($\lesssim 0.5^\circ$) FoV future missions and the Milky Way galaxy -- for the instruments with the broader fields of view, see Tab.~\ref{tab:future_missions}. For all missions we consider 1~Msec long observations of Segue~I~dSph (\extp/SFA, \extp/LAD and all \ath instruments) and the Milky Way(\extp/WFM and all \ths instruments). For the narrow-FoV instruments, we propose the ``background modeling'' approach similar to the one used for \xmm data analysis in this work. For the broad-FoV instruments, typically characterized by a complicated instrumental background, we propose the ``ON-OFF'' approach similar to the one used in this work for \spi data analysis. Where applicable the $D$-factors of the proposed observations (see Tab.~\ref{tab:future_missions}) correspond to the blank sky MW ``ON'' region located at relatively high galactic latitudes ($(\ell; b)\sim (0^\circ; 50^\circ)$) and the ``OFF'' region located further away from the GC at $(\ell; b)\sim (110^\circ; 50^\circ)$ assuming DM density profile in the MW reported in~\cite{Cautun:2019eaf}. For ``ON-OFF''-type observations the value reported in the table corresponds to the difference $D_{ON}-D_{OFF}$ of $D$-factors in considered ON and OFF regions. The $D$-factors for the background-modeling type observations performed with the narrow-FoV instruments correspond to the sum of the foreground MW and dSph's $D$-factors for the DM density profiles reported in \cite{Geringer-Sameth:2014yza} and \cite{Cautun:2019eaf}.

The sensitivities of the future missions {\extp, \ths and \ath} computed at 95\% C.L. are shown in Fig.~\ref{fig:future_constraints1} and  Fig.~\ref{fig:future_constraints2}, respectively, in comparison to \xmm and \spi limits on $\fpbh$ derived in this work.  For the broad-FoV, \spi-like future missions (\extp/WFM and \ths/XGIS) we additionally show the limits in a presence of the 1\% (un-correlated) systematics typically discussed for these missions. The limits marked ``\ths/XGIS'' correspond to the limits derived from joint observations of XGIS-S and XGIS-X instruments.

\section{Discussion}
\label{sec:discussion}
In this work, we studied the sensitivity of the current (\xmm, \spi) and future (\ath, \extp, \ths) X-ray missions for the search for the signal from {non-rotating, monochromatic mass-function} evaporating primordial black holes. Depending on the field of view of the instruments we 
selected different preferred targets 
for the observations -- a DM-dominated dSph for narrow-FoV instruments (\xmm, \extp/SFA, LAD, \ath) and the Milky Way galaxy itself for broad-FoV instruments (\extp/WFM, \ths), see Tab.~\ref{tab:future_missions}. Accordingly, we discuss different observational strategies applicable to these instruments: for the
narrow-FoV instruments -- the search for the PBH's signal on top of modeled astrophysical/instrumental background; for the broad-FoV missions typically characterized by complex for modeling and/or time-variable instrumental/astrophysical background -- ``ON-OFF'' strategy which relies on subtraction of the astrophysical and instrumental background derived from OFF-region observations. We argue that for the ``ON-OFF'' approaches the systematic uncertainty of the background measurement plays a crucial role in deriving constraints on $\fpbh$.
\begin{figure}
    \centering
    \includegraphics[width=\linewidth]{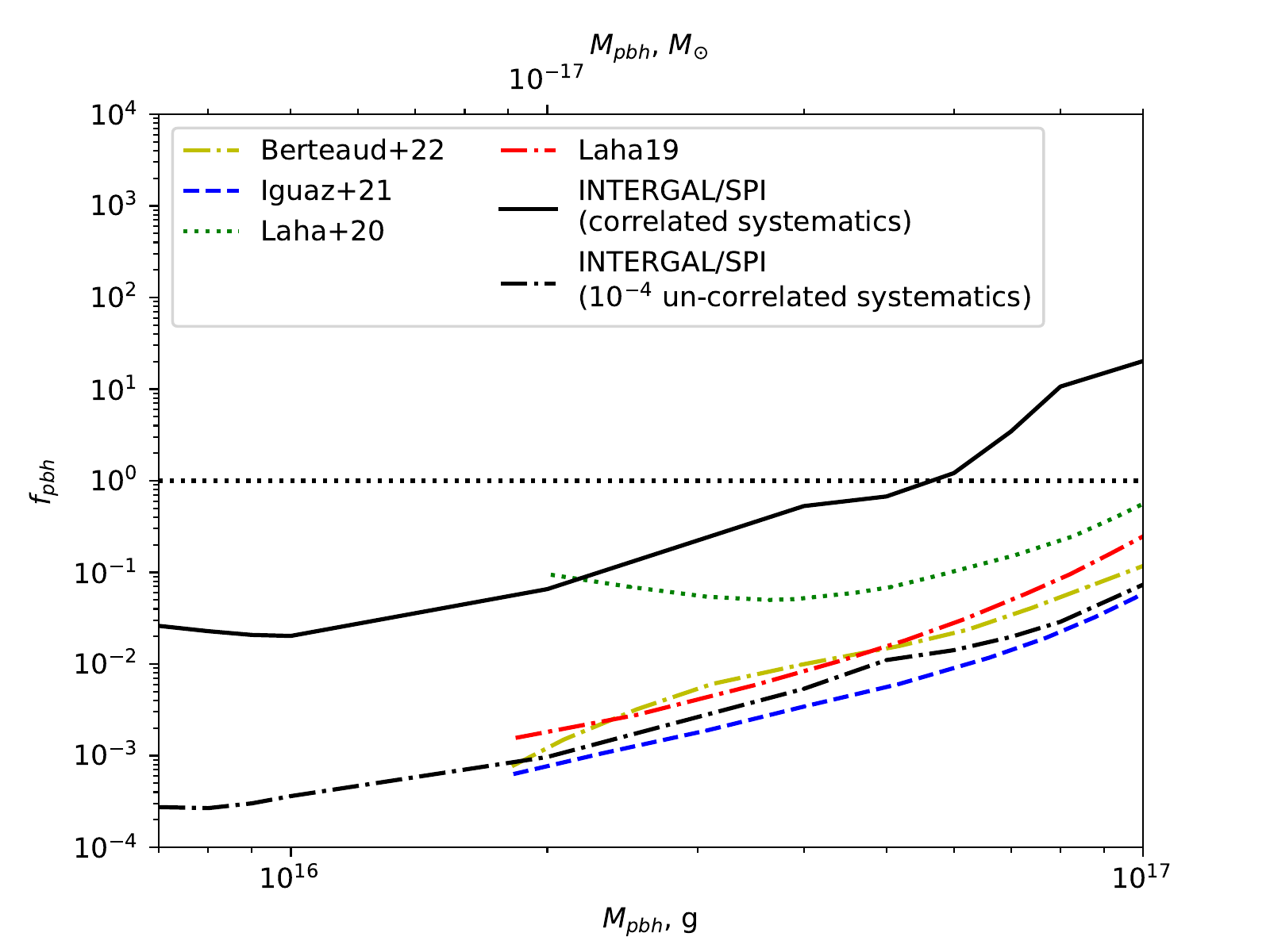}
    \caption{The constraints on $\fpbh$ derived from \spi ``ON-OFF'' observations of the MW Galaxy with 0.5\% background systematic uncertainty (solid black curve) compared to the template-based constraints from \spi data~\cite{laha20, berteaud22}, 
    constraints from 511~keV line from the GC vicinity~\cite{laha19}, %(Laha+19);
    constraints from the cosmic rays diffuse background~\cite{iguaz21}. %(Iguaz+21).
    A black dotted-dashed line illustrates the hypothetical reach of \spi sensitivity of the ``ON-OFF'' analysis similar to that presented in this work for the $10^{-4}$ systematic uncertainty. }
    \label{fig:literature_constraints}
\end{figure}

In Fig.~\ref{fig:literature_constraints} we compare the limits on $\fpbh$ derived in this paper from \spi observations of the MW galaxy to recent constraints presented in the literature~\cite{laha19, laha20, iguaz21, berteaud22}. These constraints are based on MW \spi observations accompanied by template-based modeling of the instrumental and astrophysical backgrounds~\cite{laha20, berteaud22}; constraints from the 511~keV line from the GC vicinity~\cite{laha19}; constraints from the cosmic-ray diffuse background~\cite{iguaz21}. The \spi constraints based on the ``ON-OFF'' approach presented in this work are by a factor of $\sim 50$ worse, but allow estimating the crucial role and impact of systematic uncertainties in the background modeling/subtraction not discussed for \spi in the previous works. In Fig.~\ref{fig:literature_constraints} the solid black line corresponds to the \spi constraints for 0.5\% of the ON region flux added to the statistical uncertainties of the data (correlated 0.5\% systematic case). For the illustration with the dotted-dashed curve, we show the same constraints for the hypothetical level of systematic uncertainty of 0.01\%. In this case, the derived constraints are compatible with the constraints derived in recent works. We emphasize that the accurate assessment of the systematic uncertainty is crucial for all broad-FoV missions with a time variable, complex instrumental background, and \spi in particular as an instrument with limited imaging capabilities. 

{The additional sources of uncertainties include  unknown a-priori mass-function and the average angular momentum of PBHs. 
Cosmological and astrophysical constraints on the fraction of PBH dark matter assuming an extended mass function and Kerr rotating black holes have been studied, see, for instance, Refs.~\cite{Carr:2017jsz,Kuhnel:2017pwq,Arbey:2019vqx,ray21}.
 The detailed calculations for non-monochromatic mass function and rotating PBHs are beyond the scope of this work. Following e.g.~\cite{ray21} we argue, however, that depending on the mass-function/momentum of the PBHs the derived constraints can change by an additional factor of $\sim 2$.}

The accurate treatment of systematics is additionally important since among future missions the most {prominent} perspective for the search for a signal from primordial evaporating black holes are broad-FoV instruments. Within the next decades, we can expect that the \ths/XGIS and \extp/WFM will be able to substantially improve the existing limits on $\fpbh$, see Figs.~\ref{fig:future_constraints1} and~\ref{fig:future_constraints2}. {Similar conclusions (although with different statistical methods) were derived earlier for \ths/XGIS in~\cite{ghosh22, auffinger22a} and for near-MeV missions AMEGO and GECCO in~\cite{coogan21, ray21}.} At the same time, narrow FoV, excellent energy resolution, and effective area instruments (\ath, \extp/LAD, SFA) are ideal for the search for line-like signals from decaying dark matter~\cite{we_athena,we_extp,we_theseus} are unlikely to provide %compatible
competitive limits in case the systematics for the broad-FoV instruments will be controlled at $\lesssim 1$\% level.

\section*{Acknowledgements}
The authors acknowledge support by the state of Baden-W\"urttemberg through bwHPC. DM work was supported by DLR through grant 50OR2104 and by DFG through the
grant MA 7807/2-1. The authors acknowledge fruitful discussions with J. Auffinger on the usage of the BlackHawk software. The presented work is based on observations with INTEGRAL, an ESA project with instruments and a science data center funded by ESA member states.
\input{journals.tex}
\bibliography{bibl}
\end{document}

%% file: journals.tex
% Bibliography and bibfile
\def\aj{AJ}%
          % Astronomical Journal
\def\actaa{Acta Astron.}%
          % Acta Astronomica
\def\araa{ARA\&A}%
          % Annual Review of Astron and Astrophys
\def\apj{ApJ}%
          % Astrophysical Journal
\def\apjl{ApJ}%
          % Astrophysical Journal, Letters
\def\apjs{ApJS}%
          % Astrophysical Journal, Supplement
\def\ao{Appl.~Opt.}%
          % Applied Optics
\def\apss{Ap\&SS}%
          % Astrophysics and Space Science
\def\aap{A\&A}%
          % Astronomy and Astrophysics
\def\aapr{A\&A~Rev.}%
          % Astronomy and Astrophysics Reviews
\def\aaps{A\&AS}%
          % Astronomy and Astrophysics, Supplement
\def\azh{AZh}%
          % Astronomicheskii Zhurnal
\def\baas{BAAS}%
          % Bulletin of the AAS
\def\bac{Bull. astr. Inst. Czechosl.}%
          % Bulletin of the Astronomical Institutes of Czechoslovakia
\def\caa{Chinese Astron. Astrophys.}%
          % Chinese Astronomy and Astrophysics
\def\cjaa{Chinese J. Astron. Astrophys.}%
          % Chinese Journal of Astronomy and Astrophysics
\def\icarus{Icarus}%
          % Icarus
\def\jcap{J. Cosmology Astropart. Phys.}%
          % Journal of Cosmology and Astroparticle Physics
\def\jrasc{JRASC}%
          % Journal of the RAS of Canada
\def\mnras{MNRAS}%
          % Monthly Notices of the RAS
\def\memras{MmRAS}%
          % Memoirs of the RAS
\def\na{New A}%
          % New Astronomy
\def\nar{New A Rev.}%
          % New Astronomy Review
\def\pasa{PASA}%
          % Publications of the Astron. Soc. of Australia
\def\pra{Phys.~Rev.~A}%
          % Physical Review A: General Physics
\def\prb{Phys.~Rev.~B}%
          % Physical Review B: Solid State
\def\prc{Phys.~Rev.~C}%
          % Physical Review C
\def\prd{Phys.~Rev.~D}%
          % Physical Review D
\def\pre{Phys.~Rev.~E}%
          % Physical Review E
\def\prl{Phys.~Rev.~Lett.}%
          % Physical Review Letters
\def\pasp{PASP}%
          % Publications of the ASP
\def\pasj{PASJ}%
          % Publications of the ASJ
\def\qjras{QJRAS}%
          % Quarterly Journal of the RAS
\def\rmxaa{Rev. Mexicana Astron. Astrofis.}%
          % Revista Mexicana de Astronomia y Astrofisica
\def\skytel{S\&T}%
          % Sky and Telescope
\def\solphys{Sol.~Phys.}%
          % Solar Physics
\def\sovast{Soviet~Ast.}%
          % Soviet Astronomy
\def\ssr{Space~Sci.~Rev.}%
          % Space Science Reviews
\def\zap{ZAp}%
          % Zeitschrift fuer Astrophysik
\def\nat{Nature}%
          % Nature
\def\iaucirc{IAU~Circ.}%
          % IAU Cirulars
\def\aplett{Astrophys.~Lett.}%
          % Astrophysics Letters
\def\apspr{Astrophys.~Space~Phys.~Res.}%
          % Astrophysics Space Physics Research
\def\bain{Bull.~Astron.~Inst.~Netherlands}%
          % Bulletin Astronomical Institute of the Netherlands
\def\fcp{Fund.~Cosmic~Phys.}%
          % Fundamental Cosmic Physics
\def\gca{Geochim.~Cosmochim.~Acta}%
          % Geochimica Cosmochimica Acta
\def\grl{Geophys.~Res.~Lett.}%
          % Geophysics Research Letters
\def\jcp{J.~Chem.~Phys.}%
          % Journal of Chemical Physics
\def\jgr{J.~Geophys.~Res.}%
          % Journal of Geophysics Research
\def\jqsrt{J.~Quant.~Spec.~Radiat.~Transf.}%
          % Journal of Quantitiative Spectroscopy and Radiative Trasfer
\def\memsai{Mem.~Soc.~Astron.~Italiana}%
          % Mem. Societa Astronomica Italiana
\def\nphysa{Nucl.~Phys.~A}%
          % Nuclear Physics A
\def\physrep{Phys.~Rep.}%
          % Physics Reports
\def\physscr{Phys.~Scr}%
          % Physica Scripta
\def\planss{Planet.~Space~Sci.}%
          % Planetary Space Science
\def\procspie{Proc.~SPIE}%
          % Proceedings of the SPIE
\let\astap=\aap
\let\apjlett=\apjl
\let\apjsupp=\apjs
\let\applopt=\ao